# The materials data ecosystem: materials data science and its role in data-driven materials discovery[*]

Hai-Qing Yin(尹海清)[1,2,3,†], Xue Jiang(姜雪)[1,3], Guo-Quan Liu(刘国权)[1,3], Sharon Elder[4], Bin Xu[1], Qing-Jun Zheng(郑清军)[5], and Xuan-Hui Qu(曲选辉)[1,2,6]

[1] *Collaborative Innovation Center of Steel Technology, University of Science and Technology Beijing, Beijing, China,100083*

[2] *Beijing Laboratory of Metallic Materials and Processing for Modern Transportation, University of Science and Technology Beijing, Beijing 100083, China*

[3] *Beijing Key Laboratory of Materials Genome Engineering, University of Science and Technology Beijing, Beijing 100083, China*

[4] *Computer Science and Engineering, ~~The College of Information Sciences and Technology~~, The Pennsylvania State University, University Park, Pennsylvania, 16802, USA*

[5] *Kennametal Inc., 1600 Technology Way Latrobe, PA, 15650, USA*

[6] *Institute of Advanced Materials and Technology, University of Science and Technology Beijing, Beijing, China,100083*

Since its launch in 2011, Materials Genome Initiative (MGI) has drawn the attention of researchers from across academia, government, and industry worldwide. As one of the three tools of MGI, the materials data, for the first time, emerged as an extremely significant approach in materials discovery. Data science has been applied in different disciplines as an interdisciplinary field to extract knowledge from the data. The concept of materials data science was utilized to demonstrate the data application in materials science. To explore its potential as an active research branch in the big data age, a three-tier system was put forward to define the infrastructure of data classification, curation and knowledge extraction of materials data.

## 1. Introduction

[*] Project supported by National Key R&D Program of China (2016YFB0700503)
[†] Corresponding author. E-mail: hqyin@ustb.edu.cn

As a practice of obtaining information and insight from data, data science has become a very familiar term to researchers from various disciplines, [1]. This concept was first introduced in the 1960s and it lasted for a few decades. Statistician, CF Jeff Wu, used the term again as a discipline as extension of statistics in 1996. Nowadays, massive scientific data are produced by simulations, high-throughput scientific instruments, satellites, telescopes, and so on. The availability of big data is revolutionizing how research is conducted and leads to the emergence of a new paradigm of science based on data-intensive computing and analytics. Data science is defined as the fourth paradigm for the data intensive scientific discovery, besides experimentation, theory and calculation [2]. Furthermore, the release of the Big Data R & D Initiative in 2012 accelerates the development of the data science.

Data science has come to be applied in diverse disciplines in recent years. An integrated data science pipeline is used to identify latent signals for QT-DDIs by using electrocardiogram data in electronic health records [3]. A data-driven approach was built to simulate human mobility, provide spatially models and predict the nation-wide consequences of a massive switch to electric vehicles [4]. To address the challenges from the prevailing development of big data, besides the Data Science Journal, new journals including The Journal of Data Science and Analytics (JDSA) [5] and The Data Science and Engineering Journal (DSE) [6], were launched to stimulate scientific innovation and practice in data management and data-intensive applications. In this study, we will introduce a data ecosystem to materials data and aim to organize it into a coherent portrait for the scientific study of materials data, which are related to each other, and to the materials science and engineering discipline. The materials data ecosystem is comprised of data sources and data science. There are diverse kinds of materials data sources: publications, records from a facility and computation tools, third-party data and so on, which are not covered in this text in detail. The material data science, inherently cross-functional and at the very highest level of data study, is what is investigated here.

2. **Data science in Material science and Engineering**

In 1999, John R. Rodgers introduced a new concept as materials informatics and defined it as an effective data management for new materials discovery. Somewhat later, ICME (2008) and materials Genome Initiative (MGI, 2011) attracted more attention worldwide on integrating computational capabilities, data management, and experimental techniques together [7]. Although materials databases were built in many countries with universal access to abundant scientific data, and some have become the fundamental to material computation, materials data and materials informatics [8] witnessed their first recognition when compared with computation and experimentation in materials innovation. A concept of materials data infrastructure was put forward based on the integration of Integrated Computational Materials Engineering (ICME) and materials informatics [9]. However, the diversity of materials science has yet to be exhibited. Therefore, a system which enables one to virtually express real-world materials details, as well as, data mining needs to be built.

## 3. Infrastructure of Material Data Science

Material Data Science is the data science applied in materials science and engineering, aiming to take advantage of data in discovering the nature beneath the phenomena and production. It is a new, inherently cross-disciplinary approach. Currently, the market's ever-demanding requirements push people to need to understand the whole production chain of materials. Therefore, materials researchers strive to build a linkage between the five core points, which include the chemical composition, microstructure, manufacturing, property and the performance in service.

Knowledge engineering connects data with information, knowledge and intelligence from the concept aspect. When combined with scientific disciplines, data are endowed with disciplinary meaning and the complicated correlation among data will be explicated. To promote data science into a sub-field in materials science and engineering, a three-tier infrastructure of materials data science was introduced in this case, as shown in Fig.1, showing the main procedure of materials data development, that is, data production, management and application. The first and fundamental tier is the material data system, the second tier is life cycle curation of materials data, and the third and highest tier is material data mining and deep extraction.

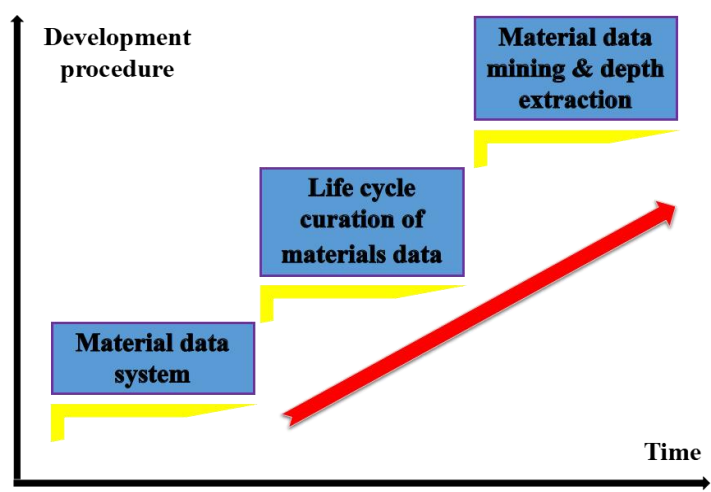

Fig. 1 Infrastructure of Material Data Science

### 3.1. The Scientific system of the material data

The generation of a single material datum, represents a kind of attribute with the characteristics of a specific material, and may have the potential to apply to a limited scope according to the material's nature and its inherent correlation. As there are several ways of classification for materials, each of which has pros and cons, the classification of materials data is definitely the same. So borrowed from the materials scientific data sharing network [10], the materials data system consists totally of 11 categories on the first level shown in Fig.2, including the material fundamental data, metal & alloy data, ceramics, organic material data, composite data, biomaterial data, information material data, energy material data, natural material data, building material data, road & transport material data. With the material data system, each single datum will be classified into a category, which will provide the following data curation steps and ultimately the data applications for the specific groups of materials. In the data system, fundamental data apparently differs from other categories of a certain material, with its primary role of providing the overall information of the single chemical elements. The crystal structure, and the properties such as the thermodynamic parameters of the binary, ternary, and combinations of elements are also pivotal contents of the fundamental data category.

The categories can be further classified into more sub-layers according to the classification system of each individual material. Furthermore, the system is also feasible as a basis of knowledge for constructing the metadata system of the materials.

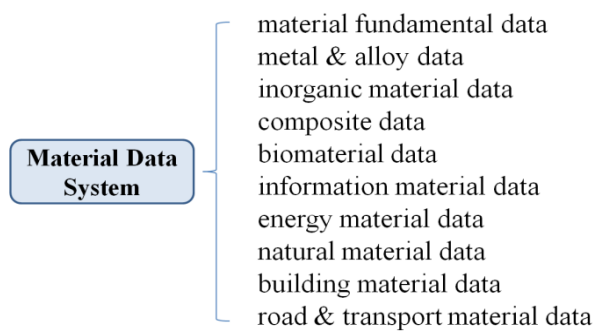

Fig. 2 The first level of material data system, borrowed from the materials scientific data sharing network. When data science is incorporated into the materials science, the material theories as well as the data standards are integrated to ensure the accurate data representation on material discipline.

### 3.2. Life cycle curation of materials data

Data, which are the product in the data era though, exhibit common characteristics of the real commodities and experience a similar development process. The life cycle of data involves production, storage, updates, management, publication, application, and finally deletion or long-term storage for reuse. The comparison of the life cycle process, shown in Fig.3, indicates the similarity between the scientific data and industrial products.

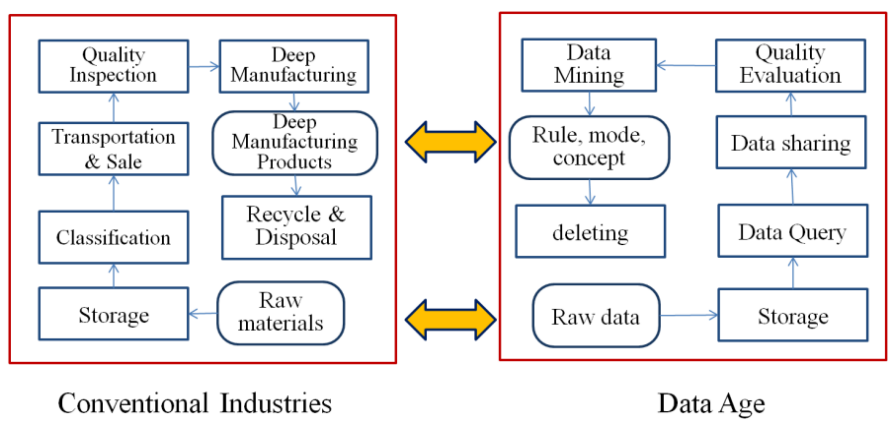

Fig. 3 Analogy of product mixes in data era and the real economy.

### 3.2.1. Description System of Materials Data

The description of scientific data is associated with storage and presentation activities. Based on the material data system, as well as, data sources from computation, experimentation, characterization and industries, we are drafting a standard for materials scientific data description, whereby the material data are divided into three groups, that is, the experimental data, the computational data and the production data. The experimental data are further divided into two sub-groups, one is the experimental data of bulk materials and the other the data on specific subjects of materials such as coatings and corrosion.

For different groups and sub-groups, the attributes being collected into the databases are different due to their inherent generation and usage, which requires comprehensive information covering everything from quantum to the macro scales in academic activities, as described in Integrated Computational Materials Engineering. As MGI describes, the goal of half-time and half-cost will be fulfilled when the endeavor focuses on an innovative reuse of the data. To meet the requirement for innovative material discovery, it is mandatory to add a detailed description of the material production process for experimental and prerequisites for computational data, which were occasionally omitted in the past. In the past, due to the limited approaches in materials research, most of the information about the data generation processes was omitted, and the reuse of the data was restricted solely to the materials performance query.

To emphasize the whole production chain of materials production and optimization, the integrity of each item of data is especially significant. The key attributes for the three groups of materials data are listed in Table. 1. It should be noted that the data from both the intermediate stages and the non-optimized samples are collected due to their potential in materials design [11] and optimization of final properties.

Table 1 Attributes of the computational, experimental and production data in the infrastructure of MGI and data mining.

| Data type | | Attributes in general (besides the data source and data colle |
|---|---|---|
| experimental data | Data Of bulk materials | ● brand / material name<br>● chemical compositions<br>● raw material information<br>● property data, testing condition, equipment & institution<br>● unstructured / descriptive file (eg. SEM, image)<br>● processing parameters |
| | Data on | ● basic material information (brand, material name, chemical compositions, property data, supplier) |

| | specific research | <ul><li>property data, testing condition, equipment & institution</li><li>unstructured / descriptive file (eg. SEM, image)</li><li>processing parameters</li></ul> |
|---|---|---|
| Computational data | | <ul><li>material name/structure</li><li>software information</li><li>model/ algorithm and its source</li><li>chemical compositions</li><li>property files / data</li><li>constraints of computation (prerequisite, initial condition, boundary)</li></ul> |
| Industrial data | | <ul><li>brand / material name</li><li>property data</li><li>description of product</li><li>company information</li></ul> |

**3.2.2.** Storage of materials data

Database, an organized collection of data, has been the typical way to define, store, update and administer data whereby the data are accessible to query and retrieval. ICSD (Inorganic crystal structure database), Paulling file, databases for thermodynamic computation and so on are those specifically used and associated with the software of first principle calculation, thermodynamics and simulation for property. Others are mostly about the properties obtained from past research and industrial activities.

Database is well developed for curation of raw materials data, while data warehouse appeared in recent a few years for data mining to store specific-topic and integrated data from one or more disparate sources. Besides the database and data warehouse, cloud storage provides a brand-new choice. Currently, cloud computing has been applied to provide Paas and Iaas service with both hardware and software equipped in some supercomputing centers and companies.

The cloud computing platform will definitely be utilized by more material data researches once the privacy and intellectual property issues in materials communities are settled.

Therefore, database, data warehouse and cloud storage are three alternative candidates for materials researchers to optimize data storage for their own data resource.

### 3.2.3. Science-Data sharing & publication

In recent years the contradiction between data sharing and proprietary interest protection becomes more obvious and is turning into a global problem to tackle. In the academic circle, which is one of the most significant sources of scientific data, the owners of data are reluctant or even reject sharing any supplemental information, as well as, the final outcomes which are essential to understand or reproduce due to protection of intellectual properties and the potential commercial values, although government funding agencies and science journals require one to do so. In material science and the like, issues related to national security on some specific topics are serious to handle and the data are prohibited to share. So difficulty to discriminate an accurate boundary impedes communication on data. Tim Austin [12] considered data citation was a choice of solution.

The combination of two widely-available methods on papers, which are identification and citation, may be one of the feasible solutions. The digital object identification (DOI) has been a common method worldwide for protection of the intellectual property of the papers in publication in recent years [13], by which the papers are uniquely labelled with a serial of numbers and letters once the registration of DOI is done. Similarly, DOI for scientific data began for geographic data in China a few years ago [14] and now a formal and detailed format for registration and citation has been established. Later a DOI system for materials data was founded in China, based on the work on the National Materials Scientific Data Sharing Network, shown in Fig 4. There are two sets of the system, coupled with each other, able to express the registered data/data set uniquely. For both systems, the data resource organization, as the university or institute mostly, tell where the data come from. The code of materials data classification, where mater is the abbreviation of term material, consists of two levels of materials data classification in chapter 3.1, with the first level shown in Fig.2. The CSTR (Science and technology resource identification) fits all the digital objects including data.

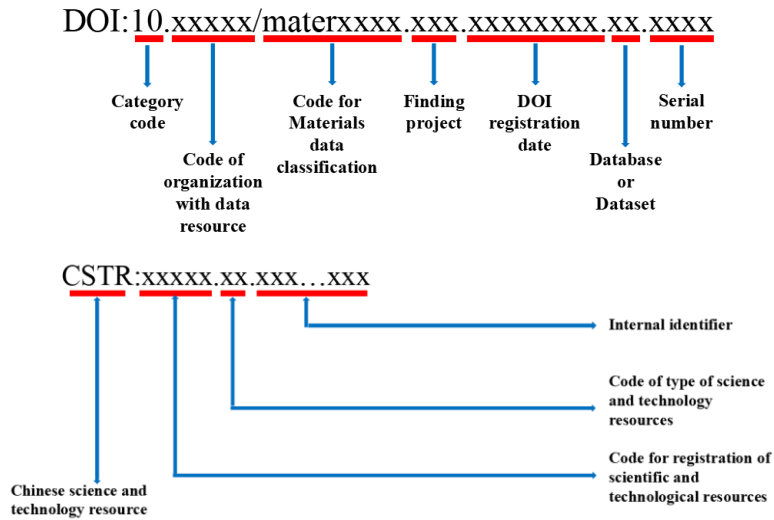

Fig. 4 The DOI and CSTR system for identifying the materials data.

Publication of scientific data is regarded as one of the ways for data sharing as well as evaluating the contributions of data collectors and DOI/CSTR provide the information of the data as a metadata for database management and information query and retrieval [15]. Sungbum Park et al produced an IS success model for evaluating the application of the DOI system and indicate that both data content including the features and information quality are significant factors to influence organizational benefits by means of perceived usefulness and user satisfaction [16]. Accordingly, DOI of materials data should be implemented at the moment when data are collected and integrated into the databases due to its high correlation to the application of databases. Unlike human health data which is pushing ahead a global coalition of data resources [17], currently, an international accessible materials data infrastructure hardly occurs, however, DOI is paving the way towards this goal.

**3.2.4.** Data transfer in Across Scale modeling and simulation

A study on across scale modeling is an interesting topic after the exploitation of multi scale modeling [18]. Smart manufacturing of materials requires across scale modeling, simulation and control by taking advantage of a combination of information with materials knowledge, where data are the fundamental element and data transfer across the scale is crucial. So the characteristics of big data in the smart manufacturing of materials are high dimension and complicated correlation rather than high volume.

Krishna Rajan pointed out that there lacks a unified way to explore patterns of behavior across correlative databases currently [19]. To bridging the gaps over the databases and across scale research activities, it is essential to understand the input and output for each scale, that is, the relevant attributes as prerequisites and boundary conditions for computation/experiments, and results in a data format. A knowledge-based understanding of the exploitation process of powder metallurgy materials is shown in Fig.5, where all the input parameters and the output are listed for scales of both the computation phase and the fabrication one. As you can see, the output does not fit the input exactly on the subsequent scale, which indicates that a comprehensive understanding of the whole chain of material design and production will be the only solution. An interface is required to bridge the gaps with input and output between two scales who have a close relationship with each other, when an across scale computation is expected.

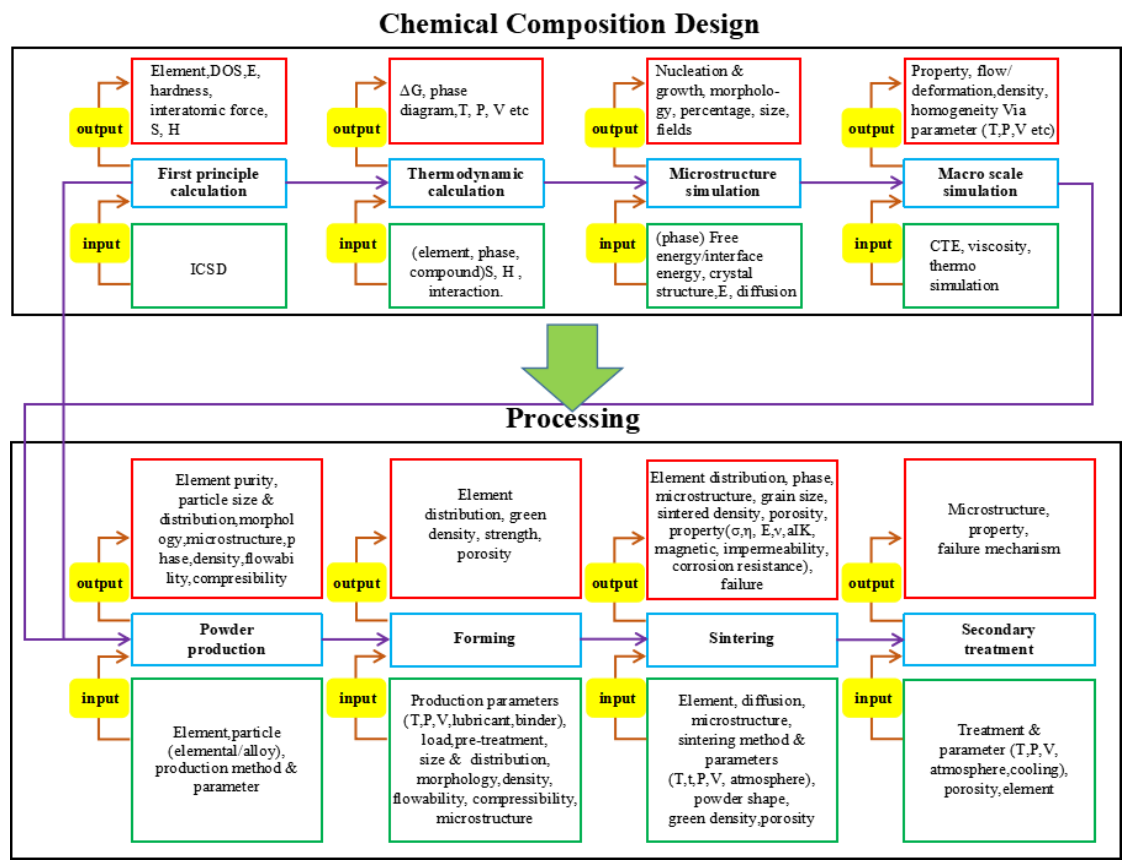

Fig.5 Computation in different scales of modeling (chemical composition design) and simulation (processing optimization) with the input parameters and output shows the feasibility of data transfer among the scales.

### 3.3. Data-driven Materials Research

Data science is regarded as the fourth paradigm for the data intensive scientific discovery, besides experimentation, theory and calculation [2]. In materials science and engineering, with the emergence of materials informatics, integrated computational material engineering, and the materials genome, materials data goes beyond the collection and integration and enters a new stage of application for the exploration and discovery of new or alternative materials, which is the core of data-driven materials research, a further step ahead of the materials by design. Material informatics turns to be a methodology of the activities for data mining and machine learning in materials science [20].

According to its functionality, the research of data mining in materials science is divided into two categories, one for creation of new materials based mainly on first principle calculation, the other for improvement of properties by optimizing the composition and processing.

In the past a few years, the breakthrough on combining MGI and data mining has emerged swiftly. Discovery of brand-new functional materials candidates, especially for clean energy storage, has been frequently reported in the Journal of Science and Nature and the term "materials code" appeared on the cover article of Journal Nature [21]. The high throughput first principle calculation (HTC) make it possible to obtain massive volumes of data, which turns to be the most abundant data resource for data mining and machine leaning[22]. Materials Project [23], AFLOWLib (Automatic Flow for Materials Discovery), OQMD (Open Quantum Materials Database), NoMaD (Novel Materials Discovery) repository, CatApp Database and CMR (Computational Materials Repository) [24] are the newly-established ab initio databases, where millions of data are integrated. By using the methods of principal component analysis, regression, neural networks and Bayesian algorithm, materials with tailored properties were discovered, such as $Ti_{50.0}Ni_{46.7}Cu_{0.8}Fe_{2.3}Pd_{0.2}$ [25,26].

Integration of materials design and optimization of processing [27] boosts investigations to solve the problems for the full work flow [28,29]. In this case, the data covers from the calculated elements to the detailed parameters in fabrication, and data mining extends the ideas of ICME to extract the semantic connections, which are central to solving tough problems of integration, cleaning, and analysis, among the attributes in experimentation and large-scale production [30,10].

4. **Prospective and challenge**

Materials data are playing a vital role in materials research. Industrial applications of material data will be a positive stimulus for the systematically establishment and implementation of materials data science on research as well as education. Smart manufacturing aims to take advantage of advanced information and manufacturing technologies to enable flexibility in physical processes, therefore Industria 4.0 enables one to apply the data in the whole work flow and the opportunity to push materials data science forward into a knowledge engineering system to realize the artificial intelligence (AI) in the materials innovation and production.

Materials data science, as a description of data science, is an interdisciplinary knowledge, which combines materials science with computer science, math as well as physics and chemistry. Collaboration is urgently needed to move one step ahead towards the requirements and goals, of which one of the core issues to overcome is the integration of material theories and knowledge with the algorithms and methods of data mining and machine learning.

**Acknowledgment**

The authors would like to thank the National Key R&D Program of China (2016YFB0700503), the National High Technology Research and Development Program of China (2015AA03420), Beijing Municipal Science and Technology Project (D161100002416001), National Natural Science Foundation of China (No.51172018), and Kennametal Inc. for the financial support.